\documentclass[nofootinbib,showpacs,preprintnumbers,amsmath,amssymb,floatfix]
{revtex4}
\usepackage{epsf}

\begin{document}

%\begin{center}

\title{Total resummaion of leading logarithms vs standard description of the Polarized DIS}

\vspace*{0.3 cm}

\author{B.I.~Ermolaev}
\affiliation{Ioffe Physico-Technical Institute, 194021
  St.Petersburg, Russia}
\author{M.~Greco}
\affiliation{Department of Physics and INFN, University Rome III,
Rome, Italy}
\author{S.I.~Troyan}
\affiliation{St.Petersburg Institute of Nuclear Physics, 188300
Gatchina, Russia}

\begin{abstract}
Total resummation  of leading logarithms of $x$ contributing to
the spin-dependent structure function $g_1$ ensures its steep rise
at small $x$.  DGLAP lacks such a resummation. Instead, the DGLAP
expressions for $g_1$ are complemented with special
phenomenological fits for the initial parton densities. The
singular factors $x^{-\alpha}$ in the fits mimic the resummation
and also ensure the steep (power-like) rise of $g_1$ at the
small-$x$ region. Furthermore, DGLAP by definition cannot describe
the region of small $Q^2$ whereas our approach can do it.
\end{abstract}


\maketitle

\section{Introduction}

The Standard Approach (SA) for theoretical investigation of  DIS
structure function $g_1(x, Q^2)$ involves DGLAP\cite{dglap} and
Standard fits\cite{fits} for the initial parton densities. In the
SA framework, $g_1^{DGLAP}$ is a convolution of the coefficient
functions $C_{DGLAP}$ and evolved (with respect to $Q^2$) parton
distributions which are also expressed as a convolution of the
splitting functions $P_{DGLAP}$ and initial parton densities. The
latter are found from experimental data at large $x$, $x \sim 1$
and  $Q^2 \sim 1$~GeV$^2$. As a result, SA accounts for the $Q^2$
-evolution through the DGLAP evolution equations whereas the $x$
-evolution is accounted for through the fits which are found from
phenomenological considerations.  The reason for such asymmetric
treating the $Q^2$ and $x$ -evolutions in SA is that  DGLAP was
originally constructed for operating at large $x$ where $x$-
contributions from higher loops were small and could be neglected.
In other words, the $x$ -evolution can be neglected at large $x$.
However, in the small-$x$ region the situation looks opposite:
logarithms of $x$ are becoming quite sizable and should be
accounted to all orders in $\alpha_s$. The total resummation of
leading logarithms of $x$ was first done in Refs.~\cite{ber} in
the double-logarithmic approximation so that $\alpha_s$ was kept
fixed at an unknown scale and later in Refs.~\cite{egt2} where the
running $\alpha_s$ effects were accounted for. Contrary to DGLAP
where
\begin{equation}\label{dglapparam}
\alpha_s^{DGLAP} = \alpha_s(Q^2),
\end{equation}
Ref.~\cite{egt2} used the parametrization of $\alpha_s$ suggested
in Ref.~\cite{egt1} because the DGLAP parametrization of
Eq.~(\ref{dglapparam}) cannot be used at small $x$.   The
parametrization of Ref.~\cite{egt1} is universally good for both
small $x$ and large $x$.  It converge to the DGLAP-
parametrization at large $x$ but differs from it at small $x$.

Nevertheless, it is known that, despite DGLAP lacks the total
resummation of $\ln x$, it successfully operates at $x \ll 1$. As
a result, the common opinion was formed that not only the total
resummation of DL contributions in Refs.~\cite{ber} but also the
much more accurate calculations performed in Refs.~\cite{egt2}
should be out of use at available $x$ and might be of some
importance in a distant future at extremely small $x$. In
Ref.~\cite{egt3} we argued against such a point of view and
explained why SA can be so successful at small $x$: in order to be
able to describe the available experimental data, SA uses the
singular fits of Refs.~\cite{fits} for the initial parton
densities. Singular factors $x^{-a}$ in the fits mimic the total
resummaton of Refs.~\cite{egt2}. Using the results of
Ref~\cite{egt2} allows to simplify the rather sophisticated
structure of the standard fits.

Another essential difference between SA and our description of
$g_1$ is the obvious fact that DGLAP works at the kinematic
regions of large $Q^2$ whereas our approach is valid for large and
small $Q^2$. The latter is important in particular for theoretical
explanation of the COMPASS collaboration results. In
Ref.~\cite{egtsmq} we showed that $g_1$ practically does not
depend on $x$ at small $x$, even at $x \ll 1$. Instead, it depends
on the total invariant energy $2pq$. Experimental investigation of
this dependence is extremely interesting because according to our
results $g_1$, being positive at small $2pq$, can turn negative at
greater values of this variable. The position of the turning point
is sensitive to the ratio between the initial quark and gluon
densities, so its experimental detection  would enable to estimate
this ratio.

\section{Difference between DGLAP and our approach}

 In DGLAP, $g_1$ is expressed through convolutions of the coefficient functions
and evolved parton distributions. As convolutions look simpler in
terms of integral transforms, it is convenient to represent $g_1$
in the form of the Mellin integral. For example, the non-singlet
component  of $g_1$ can be represented as follows:

\begin{equation}
\label{fdglapmellin} g^{NS}_{1~DGLAP}(x, Q^2) = (e^2_q/2)
\int_{-\imath \infty}^{\imath \infty} \frac{d \omega}{2\imath
\pi}(1/x)^{\omega} C_{DGLAP}(\omega) \delta q(\omega) \exp \Big[
\int_{\mu^2}^{Q^2} \frac{d k^2_{\perp}}{k^2_{\perp}}
\gamma_{DGLAP}(\omega, \alpha_s(k^2_{\perp}))\Big]
\end{equation}
with $C_{DGLAP}(\omega)$ being the non-singlet coefficient
functions, $\gamma_{DGLAP}(\omega,  \alpha_s)$ the non-singlet
anomalous dimensions and $\delta q(\omega)$ the initial
non-singlet quark densities in the  Mellin (momentum) space. The
expression for the singlet $g_1$ is similar, though more involved.
Both $\gamma_{DGLAP}$ and $C_{DGLAP}$  are known in first two
orders of the perturbative QCD. Technically, it is simpler to
calculate them at integer values of $\omega = n$. In this case,
the
 integrand of Eq.~(\ref{fdglapmellin})  is called the $n$-th momentum of
 $g^{NS}_1$. When the moments for different $n$ are known, $g^{NS}$ at arbitrary values of $\omega$
 is obtained with interpolation of the moments.
 Expressions for the initial quark densities are defined from phenomenological consideration,
 with fitting experimental data at $x \sim 1$.  Eq.~(\ref{fdglapmellin})  shows that $\gamma_{DGLAP}$
 govern the $Q^2$- evolution whereas  $C_{DGLAP}$ evolve
 $\delta q(\omega)$
 in the $x$-space from $x \sim 1$ into the small$x$ region.
 When, at the $x$-space, the initial parton distributions $\delta
q(x)$ are regular in $x$, i.e. do not $\to \infty$ when $x \to 0$,
the small-$x$ asymptotics of $g_{1~DGLAP}$ is given by the
well-known expression:
\begin{equation}\label{dglapas}
 g^{NS}_{1~DGLAP},~~g^{S}_{1~DGLAP} \sim \exp \Big[ \sqrt{\ln (1/x) \ln
 \Big( \ln(Q^2/\mu^2)/\ln (\mu^2/
 \Lambda^2_{QCD})\Big)}~\Big].
\end{equation} On the contrary, when the total resummation of the double-logarithms (DL) and single-
logarithms of $x$ is done\cite{egt1}, the Mellin representation
for $g_1^{NS}$  is
\begin{equation}
\label{gnsint} g_1^{NS}(x, Q^2) = (e^2_q/2) \int_{-\imath
\infty}^{\imath \infty} \frac{d \omega}{2\pi\imath }(1/x)^{\omega}
C_{NS}(\omega) \delta q(\omega) \exp\big( H_{NS}(\omega)
\ln(Q^2/\mu^2)\big)~,
\end{equation}
with new coefficient functions  $C_{NS}$,
\begin{equation}
\label{cns} C_{NS}(\omega) =\frac{\omega}{\omega -
H_{NS}^{(\pm)}(\omega)}
\end{equation}
and anomalous dimensions $H_{NS}$,
\begin{equation}
\label{hns} H_{NS} = (1/2) \Big[\omega - \sqrt{\omega^2 -
B(\omega)} \Big]
\end{equation}
where
\begin{equation}
\label{b} B(\omega) = (4\pi C_F (1 +  \omega/2) A(\omega) +
D(\omega))/ (2 \pi^2)~.
\end{equation}
 $ D(\omega)$ and $A(\omega)$ in Eq.~(\ref{b}) are
expressed in terms of  $\rho = \ln(1/x)$, $\eta =
\ln(\mu^2/\Lambda^2_{QCD})$, $b = (33 - 2n_f)/12\pi$ and the color
factors
 $C_F = 4/3$, $N = 3$:

\begin{equation}
\label{d} D(\omega) = \frac{2C_F}{b^2 N} \int_0^{\infty} d \rho
e^{-\omega \rho} \ln \big( \frac{\rho + \eta}{\eta}\big) \Big[
\frac{\rho + \eta}{(\rho + \eta)^2 + \pi^2} \mp
\frac{1}{\eta}\Big] ~,
\end{equation}

\begin{equation}
\label{a} A(\omega) = \frac{1}{b} \Big[\frac{\eta}{\eta^2 + \pi^2}
- \int_0^{\infty} \frac{d \rho e^{-\omega \rho}}{(\rho + \eta)^2 +
\pi^2} \Big].
\end{equation}
$H_{S}$  and $C_{NS}$ account for DL and SL contributions to all
orders in $\alpha_s$. When $x \to 0$,
\begin{equation}
\label{gnsas}g_1^{NS} \sim \big( x^2/Q^2\big)^{\Delta_{NS}/2},~
g_1^{S} \sim \big( x^2/Q^2\big)^{\Delta_{S}/2}
\end{equation}
where the non-singlet and singlet intercepts are $\Delta_{NS} =
0.42,~\Delta_{S} = 0.86$. The $x$- behavior of Eq.~(\ref{gnsas})
is much steeper than the one of Eq.~(\ref{dglapas}). Obviously,
the total resummation of logarithms of $x$ leads to the faster
growth of $g_1$ when $x$ decreasing compared to the one predicted
by DGLAP, providing the input $\delta q$ in
Eq.~(\ref{fdglapmellin}) is a regular function of $\omega$ at
$\omega \to 0$.

\section{Structure of the standard DGLAP fits}
Although there are different fits for  $\delta q(x)$ in
literature, all available fits include  both  regular and singular
factors when $x \to 0$. For example, the typical expression is
\begin{equation}
\label{fita} \delta q(x) = N \eta x^{- \alpha} \Big[(1
-x)^{\beta}(1 + \gamma x^{\delta})\Big],
\end{equation}
with $N,~\eta$ being a normalization, $\alpha = 0.576$, $\beta =
2.67$, $\gamma = 34.36$ and $\delta = 0.75$. In the $\omega$
-space Eq.~(\ref{fita}) is a sum of pole contributions:
\begin{equation}
\label{fitaomega} \delta q(\omega) = N \eta \Big[ (\omega -
\alpha)^{-1} + \sum m_k (\omega + \lambda_k)^{-1}\Big],
\end{equation}
with $\lambda_{k} > 0$, so that the first term in
Eq.~(\ref{fitaomega}) corresponds to the singular factor
$x^{-\alpha}$ of Eq.~(\ref{fita}). When the fit Eq.~(\ref{fita})
is  substituted in Eq.~(\ref{fdglapmellin}),  the singular factor
$x^{-\alpha}$ affects the small -$x$ behavior of $g_1$ and changes
its asymptotics Eq.~(\ref{dglapas}) for $g_1$ for the Regge
asymptotics. Indeed, the small- $x$ asymptotics is governed by the
leading singularity $\omega = \alpha$, so
\begin{equation}\label{asdglap}
g_{1~DGLAP} \sim C(\alpha)(1/x)^{\alpha}\Big((\ln(Q^2/\Lambda^2))/
(\ln(\mu^2/\Lambda^2))\Big)^{\gamma(\alpha)}.
\end{equation}
 Obviously,  the actual DGLAP asympotics of Eq.~(\ref{asdglap}) is of the Regge type,
 it differs a lot from the conventional DGLAP asympotics of
 Eq.~(\ref{dglapas}) and
 looks similar to
 our asymptotics given by Eq.~(\ref{gnsas}):  incorporating the singular
factors into DGLAP fits ensures the steep rise of $g_1^{DGLAP}$ at
small $x$ and thereby leads to the success of DGLAP at small $x$.
Ref.~\cite{egt3} demonstrates that without the singular factor
$x^{-\alpha}$ in the fit of Eq.~(\ref{fita}), DGLAP would not be
able to operate successfully at $x \leq 0.05$. In other words, the
singular factors in DGLAP fits
 mimic the total resummation of logarithms of $x$ of
Eqs.~(\ref{gnsint},\ref{gnsas}).  Although both (\ref{asdglap})
and (\ref{gnsas}) predict the Regge asymptotics for $g_1$, there
is a certain difference between them: Eq.~(\ref{asdglap}) predicts
that the intercept of $g_1^{NS}$ should be $\alpha = 0.57$.  As
$\alpha$ is greater than the non-singlet intercept $ \Delta_{NS} =
0.42$, the non-singlet $g_1^{DGLAP}$ grows, when $x \to 0$, faster
than our predictions. However, such a rise is too steep. It
contradicts the results obtained in Refs.~\cite{egt2} and
confirmed by several groups fitting HERMES data. Usually, the
DGLAP equations for the non-singlets are written in the $x$-space
as convolutions of splitting functions $P_{qq}$ with evolved
parton distributions $\Delta q$ and  the latter are written as
another convolution:

\begin{equation}\label{evol}
  \Delta q(x)  = C_q(x,y)\otimes \delta q(y) ,
\end{equation}
with $C_q$ being the coefficient function. Written in this way,
$\Delta q$ is sometimes  believed to be less singular than $\delta
q$ because of the evolution. However applying the Mellin transform
to Eq.~(\ref{evol}) immediately disproves it.

\section{$g_1$ at small $x$ and small $Q^2$}

The COMPASS experiment measures the singlet $g_1$ at $x \sim
10^{-3}$ and $Q^2 \ll 1$~GeV${^2}$, i.e. in the kinematic region
where it is impossible to use DGLAP. Although formulae for singlet
and non-singlet $g_1$ are different, with formulae for the singlet
being much more complicated, we can explain the essence of our
approach, using Eq.~(\ref{gnsint}) for the non-singlet.

\begin{figure}
\begin{center}
\begin{picture}(240,140)
\put(0,0){
\epsfbox{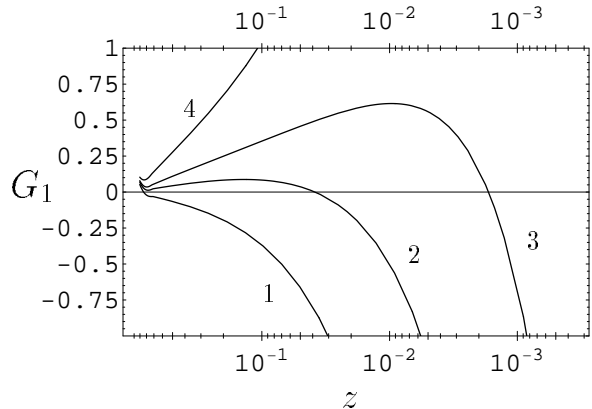}}
\end{picture}
\end{center}
\caption{$G_1$ evolution with decreasing $z=\mu^2/2(pq)$ for
different values of ratio $r= \delta g/\delta q$: curve 1 - for
$r=0$, curve 2 - for $r= -5$~, curve~3 -for $r = -8$ and curve~4
-for $r = -15$.} \label{fig1}
\end{figure}

In the COMPASS experiment $Q^2 \ll \mu^2$. The expression for
$g_1$ at such small $Q^2$ with logarithmic accuracy is given by
Eq.~(\ref{gnsint}) where the $Q^2$ -dependence is dropped and $x$
is replaced by $\mu^2/2pq$, so

\begin{equation}
\label{gnsq0} g_1^{NS}(x, Q^2 \lesssim \mu^2) = (e^2_q/2)
\int_{-\imath \infty}^{\imath \infty} \frac{d \omega}{2\pi\imath
}(\mu^2/2pq)^{\omega} C_{NS}^q(\omega) \delta q(\omega)  ~.
\end{equation}
The expression for the $g_1$ singlet looks similar, though $e_q^2$
should be replaced by the averaged charge $<e^2_q>$ and
$C_{NS}(\omega) \delta q(\omega)$ should be replaced by the sum
\begin{equation}\label{cs}
C_{S}^q(\omega) \delta q(\omega) + C_{S}^g(\omega) \delta
g(\omega)
\end{equation}
 so that $\delta q(\omega)$ and $\delta g(\omega)$ are
the initial quark and gluon densities respectively and
$C_{S}^{q,g}$ are the singlet coefficient functions. Explicit
expressions for  $C_{S}^{q,g}$ are given in Ref.~\cite{egtsmq}.
The standard fits for $\delta q$ and $\delta g$ contain singular
factors $\sim x^{-a}$ which mimic the total resummation of leading
logarithms of $x$. Such a resummation leads to the expressions for
the coefficient functions different from the DGLAP ones. After
that the singular factors in the fits can be dropped and the
initial parton densities can be approximated by constants:
\begin{equation}\label{deltaqg}
\delta q \approx N_q ~~~~\delta g \approx N_g~,
\end{equation}
so, one can write
\begin{equation}\label{g1num}
g_1(Q^2 \ll \mu^2) \approx (<e^2_q>/2) N_q G_1(z)
\end{equation}
with
\begin{equation}
\label{g1q0} G_1 =  \int_{-\imath \infty}^{\imath \infty} \frac{d
\omega}{2\pi\imath }(1/z)^{\omega} \big[C_S^q + (N_g/N_q C_g)
\big] ~
\end{equation}
where $z = \mu^2/2pq$. Obviously, $G_1$ depends on the ratio
$N_g/N_q$. The results for different values of the ratio
$r=N_g/N_q$, $G_1$ are plotted in Fig.~\ref{fig1}. When the  gluon
density is neglected, i.e. $N_g = 0$ (curve~1), $G_1$ being
positive at $x \sim 1$, is getting negative very soon, at $z <
0.5$ and falls fast with decreasing $z$. When $N_g/N_q = -5$
(curve~2), $G_1$ remains positive and not large until $z \sim
10^{-1}$, turns negative at $z \sim 0.03$ and falls afterwards
rapidly with decreasing $z$ . This turning point where $G_1$
changes  its sign is very sensitive to the magnitude of the ratio
$r$~. For instance, at $N_g/N_q = -8$ (curve~3), $G_1$ passes
through zero at $z \sim 10^{-3}$. When $N_g/N_q < -10$, $G_1$ is
positive at any experimentally reachable $z$ (curve~4)~.
Therefore, the
 experimental measurement of the turning point
would allow to draw conclusions on the interplay between the
initial quark and gluon densities.
%%%%%%%%%%%%%%%%%%%%%%%%%%%%%%%%
%%%%%%%%%%%%%%%%%%%%%%%%%%%%%%%%

\section{Conclusion}

Comparison of Eqs.~(\ref{dglapas}) and (\ref{asdglap}) shows
explicitly that the singular factor $x^{-\alpha}$ in the
Eq.~(\ref{fita}) for the initial quark density converts the
exponential DGLAP-asympotics into the Regge one. On the other
hand, comparison of Eqs.~(\ref{gnsas}) and (\ref{asdglap})
demonstrates that the singular factors in the DGLAP fits mimic the
total resummation of logarithms of $x$. These factors can be
dropped when the total resummation of logarithms of $x$ performed
in Ref.~\cite{egt2} is taken into account. The remaining, regular
$x$-terms of the DGLAP fits (the terms in squared brackets in
Eq.~(\ref{fita})) can obviously be simplified or even dropped at
small $x$ so that the rather complicated DGLAP fits can be
replaced by constants. It immediately leads to an interesting
conclusion: the DGLAP fits for $\delta q$ have been commonly
believed to represent non-perturbative QCD effects but they
actually mimic the contributions of the perturbative QCD, so the
whole impact of the non-perturbative QCD on $g_1$ at small $x$ is
not large and can be approximated by normalization constants. We
have used the latter for studying the $g_1$ singlet at small $Q^2$
because this kinematic is presently investigated by the COMPASS
collaboration. It turns out that $g_1$ in this region depends on
$z = \mu^2/2pq$ only and practically does not depend on $x$.
Numerical calculations show that the sign of $g_1$ is positive at
$z$ close to 1 and can remain positive or become negative at
smaller $z$, depending on the ratio between $\delta g$ and $\delta
q$. It is plotted in Fig.~1 for different values of $\delta
g/\delta q$. Fig.~1 demonstrates that the position of the sign
change point is sensitive to the ratio $\delta g/\delta q$, so the
experimental measurement of this point would enable to estimate
the impact of $\delta g$.

\end{document}